\documentclass[10pt,conference,a4paper]{IEEEtran}
\usepackage{times,amsmath,epsfig}
\title{Blind Adaptive MIMO
Receivers for CDMA Systems with Space-Time Block-Codes and Low-Cost
Algorithms}
\author{R. C. de Lamare \dag and R. Sampaio-Neto  \ddag \\
\fontsize{10}{10}\selectfont\itshape \dag Communication Research Group, Department of Electronics, The University of York, UK \\
 \ddag CETUC/Pontifical Catholic University of Rio de Janeiro (PUC-RIO), Brazil \\
\fontsize{9}{9}\selectfont\ttfamily\upshape E-mails:
rcdl500@ohm.york.ac.uk, raimundo@cetuc.puc-rio.br }

\begin{document}
\maketitle \thispagestyle{empty}

\begin{abstract} \small In this paper we present
low-complexity blind multi-input multi-output (MIMO) adaptive
linear multiuser receivers for direct sequence code division
multiple access (DS-CDMA) systems using multiple transmit antennas
and space-time block codes (STBC) in multipath channels. A
space-time code-constrained constant modulus (CCM) design
criterion based on constrained optimization techniques and
low-complexity stochastic gradient (SG) adaptive algorithms are
developed for estimating the parameters of the space-time linear
receivers. The receivers are designed by exploiting the unique
structure imposed by both spreading codes and STBC. A blind
space-time channel estimation scheme for STBC systems based on a
subspace approach is also proposed along with an efficient SG
algorithm. Simulation results for a downlink scenario assess the
receiver structures and algorithms and show that the proposed
schemes achieve excellent performance, outperforming existing
methods.
\end{abstract}

\section{Introduction}

The high demand for performance and capacity in wireless networks
has motivated the development of a large number of signal
processing and communications techniques for utilizing these
resources efficiently. Recent results on information theory have
shown that it is possible to achieve high spectral efficiency
\cite{foschini} and to make wireless links more reliable
\cite{alamouti,tarokh1} through the deployment of multiple
antennas at both transmitter and receiver. Space-time coding (STC)
is a set of techniques that provides an effective way of
exploiting spatial and temporal transmit diversity, resulting in
more reliable wireless links \cite{alamouti,tarokh1}. Because it
is more cost-effective to deploy multiple antennas at the base
station rather than at the mobile terminal, several space-time
coding techniques such as space-time block codes (STBC)
\cite{alamouti,tarokh1} and differential space-time codes
\cite{hugues}, have been recently investigated for improving the
quality of wireless downlink connections. Among these techniques,
STBC has become very popular as it offers maximum diversity gain
based on linear processing at the receiver and due to the
simplicity in which it can be adopted in several communications
technologies such as the widely adopted CDMA systems.

In CDMA systems, the designer usually has to deal with multiple
access interference (MAI), which arises due to lack of orthogonality
between the user signals at the receiver, and intersymbol
interference (ISI), which is a result of multiple propagation paths.
Interference suppression in multi-antenna systems is more
challenging than in single-antenna ones because the former is
subject to self-interference. The problem of receiver design for
DS-CDMA systems using multiple transmit antennas and STBC has been
considered in several recent works \cite{huang}-\cite{reynolds}.
Maximum likelihood solutions \cite{huang} that consider both STBC
and multiuser systems are very complex and this calls for suboptimal
and low-complexity solutions. Since original STBCs were originally
designed for flat channels, they may experience performance
degradation under multipath, where their diversity gain capabilities
can be considerably reduced. In order to design CDMA systems with
STBC in multipath scenarios an interesting and promising approach is
to combine the transmit streams separation and interference
suppression tasks into one single processing stage. This enables the
use of low-complexity multiuser receivers and design algorithms.

In this work we present blind adaptive linear MIMO receivers based
on the code-constrained constant modulus (CCM) design for DS-CDMA
systems using multiple transmit antennas and STBC in multipath
channels. We consider a multi-antenna system with STBC and exploit
the unique code structure of space-time coding to derive efficient
blind receivers based on the CCM design and develop low-complexity
stochastic gradient (SG) algorithms. To blindly estimate the
channel, we describe a subspace approach that exploits the STBC
structure present in the received signal and derive an adaptive SG
channel estimator. The only requirement for the receivers is the
knowledge of the signature sequences for the desired user.

This paper is organized as follows. Section II briefly describes the
space-time DS-CDMA communication system model. The space-time
linearly constrained receivers based on the CCM design criterion are
presented in Section III. A subspace framework for blind channel
estimation of DS-CDMA signals with STBC is presented in Section IV.
Section V is devoted to the derivation of adaptive SG algorithms for
both receiver and channel parameter estimation. Section VI presents
and discusses the simulation results and Section VII gives the
concluding remarks of this work.

\section{Space-time DS-CDMA system model}

Let us consider the downlink of a symbol synchronous QPSK DS-CDMA
system with $K$ users, $N$ chips per symbol, $N_{t}$ antennas at the
transmitter, $N_r$ antennas at the receiver and $L_{p}$ propagation
paths. For simplicity, we assume that the transmitter (Tx) employs
only $N_t=2$ antennas and adopts Alamouti's STBC scheme
\cite{alamouti}, even though other STBC can also be deployed.
According to this scheme, depicted in Fig. 1, during the $(2i-1)th$
symbol interval, where $i$ is an integer, two symbols $b_{k}(2i-1)$
and $b_{k}(2i)$ drawn from the same constellation are transmitted
from Tx1 and Tx2, respectively, and during the next symbol interval,
$-b_{k}(2i)$ and $b_{k}^*(2i)$ are transmitted from Tx1 and Tx2,
respectively. In addition, each user is assigned a different
spreading code for each Tx, which may be constructed from a single
spreading code ${\bf s}_k$ as $[{\bf s}_k^T, {\bf 0}^T]^T$ and
$[{\bf 0}^T, {\bf s}_k^T ]^T$, respectively, a scheme proposed for
UMTS W-CDMA or $[{\bf s}_k^T, {\bf s}_k^T]^T$ and $[{\bf s}^T_k,
-{\bf s}_k^T ]^T$, respectively, an approach adopted in the IS-2000
standard \cite{hochwald}.

\begin{figure}[!htb]
\begin{center}
\def\epsfsize#1#2{1.1\columnwidth}
\epsfbox{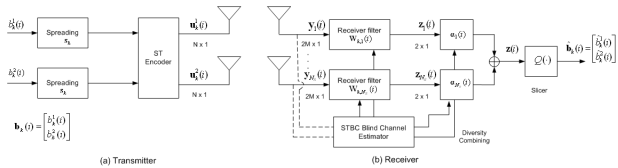} \vspace{-2.5em}\caption{Proposed space-time
system: schematic of the $k$th user (a) Transmitter and (b) Receiver
.}
\end{center}
\end{figure}

In this context, the baseband signal transmitted by the $k$-th
active user to the base station is given by
\begin{equation}
x_{k}^{1}(t)=A_{k}\sum_{i=1}^{P}b_{k}^{1}(i)s_{k}^{1}(t-2iT) -
b_{k}^{2}(i)s_{k}^{2}(t-(2i-1)T)
\end{equation}
\begin{equation}
x_{k}^{2}(t)=A_{k}\sum_{i=1}^{P}b_{k}^{2}(i)s_{k}^{2}(t-2iT) +
b_{k}^{1}(i)s_{k}^{1}(t-(2i-1)T)
\end{equation}
where $P$ is the packet length, $b_{k}(i) \in \{\pm 1 \pm j \}$ is
the $i$th symbol for user $k$ with $j^{2}=-1$, the real valued
spreading waveform and the amplitude associated with user $k$ are
$s_{k}^{1,2}(t)$ and $A_{k}$, respectively. The spreading waveforms
are expressed by $s_{k}^{1,2}(t) = \sum_{i=1}^{N} a_{k}(i)
\phi(t-iT_{c})$, where $a_{k}(i)\in \{\pm1/\sqrt{N} \}$, $\phi(t)$
is the chip waverform, $T_{c}$ is the chip duration and $N=T/T_{c}$
is the processing gain. Assuming that the receiver is synchronised
with the main path, the coherently demodulated composite received
signal for receive antenna $m$ is
\begin{equation}
\begin{split}
r_m(t) & = \sum_{k=1}^{K}\sum_{l=0}^{L_{p}-1}
h_{l,m}^1(t)x_{k,1}(t-\tau_{l,m}^1) \\ & \quad +
h_{l,m}^2(t)x_{k,2}(t-\tau_{l,m}^2) + n(t)
\end{split}
\end{equation}
where $h_{l,m}^{1,2}(t)$ and $\tau_{l,m}^{1,2}$ are, respectively,
the channel coefficients for the $m$th receive antenna, transmit
antenna $n_t=1,2$ and their corresponding delay associated with the
$l$-th path. Assuming that $\tau_{l,m}^{1,2} = lT_{c}$,
$h_{l,m}^{1,2}(i) = h_{l,m}^{1,2}(iT_{c})$, the channel is constant
during two symbol intervals and the spreading codes are repeated
from symbol to symbol, the received signal $ r(t)$ at antenna $m$
after filtering by a chip-pulse matched filter and sampled at chip
rate over two consecutive symbols yields the $M$-dimensional
received vectors
\begin{equation}
\begin{split}
{\bf r}(2i-1) & = \sum_{k=1}^{K} A_{k}b_{k}(2i-1) {\bf C}_{k}^1{\bf
h}_{m}^1+    A_{k}b_{k}(2i) {\bf C}_{k}^2{\bf h}_{m}^2 \\ & \quad +
\boldsymbol{\eta}_{k,m}(2i-1) + {\bf n}_m(2i-1)
\\ {\bf r}(2i) & = \sum_{k=1}^{K} A_{k}b_{k}^*(2i-1) {\bf C}_{k}^2{\bf
h}_{m}^2 -     A_{k}b_{k}^*(2i) {\bf C}_{k}^1{\bf
h}_{m}^1 \\ & \quad  + \boldsymbol{\eta}_{k,m}(2i) + {\bf n}_m(2i) \\
&  i=1,~\ldots,P, ~~~~~~~~ m=1,~\ldots, N_r
\end{split}
\end{equation}
where $M=N+L_{p}-1$, ${\bf n}_m(i) = [n_{1}(i)
~\ldots~n_{M}(i)]^{T}$ is the complex Gaussian noise vector with
$E[{\bf n}_m(i){\bf n}^{H}_m(i)] = \sigma^{2}{\bf I}$, where
$(.)^{T}$ and $(.)^{H}$ denote transpose and Hermitian transpose,
respectively, $E[.]$ stands for ensemble average, the amplitude of
user $k$ is $A_{k}$, the channel vector for users transmitted at Tx
$n_t$ ($n_t=1,2$) and received at Rx $m$ is ${\bf h}_{m}^{n_t} =
[h_{m,0}^{n_t} \ldots h_{m,L_{p}-1}^{n_t}]^{T}$,
$\boldsymbol{\eta}_{m}(i)$ is the intersymbol interference at Rx $m$
and the $M \times L_{p}$ convolution matrix ${\bf C}_{k}^{n_t}$
contains one-chip shifted versions of the signature sequence for
user $k$ and transmit antenna $n_t$ given by ${\bf s}_{k} =
[a_{k}(1) \ldots a_{k}(N)]^{T}$. Let us now organize the received
data in (4) into a single received vector within the $i$th symbol
interval at the $m$th receive antenna (Rx) as described by
\begin{equation}
\begin{split}
{\bf y}_{m}(i)  & = \sum_{k=1}^{K} A_{k} b_{k}(2i-1)
\boldsymbol{\mathcal{C}}_k \boldsymbol{\mathcal{H}}_m(i) +
A_{k}b_{k}(2i) \bar{\boldsymbol{\mathcal{C}}}_k
\boldsymbol{\mathcal{H}}^*_m(i)  \\ & \quad +
\boldsymbol{\eta}_k(i) + {\bf n}(i) \\
& = \sum_{k=1}^{K} {\bf x}_{k}(i) + \bar{\bf x}_{k}(i)  +
\boldsymbol{\eta}_k(i) + {\bf n}(i) \\
\end{split}
\end{equation}
where
\begin{equation}
\boldsymbol{\mathcal{C}}_k = \left[\begin{array}{c c}
{\bf C}_k^1  & {\bf 0}  \\
{\bf 0} & {\bf C}_k^2  \\
 \end{array}\right], \bar{\boldsymbol{\mathcal{C}}}_k = \left[\begin{array}{c c}
{\bf 0} & {\bf C}_k^2  \\
-{\bf C}_k^1 & {\bf 0}   \\
 \end{array}\right], ~{\bf C}_k^{1,2} \in \boldsymbol{\mathcal{R}}^M
\end{equation}
\begin{equation}
\boldsymbol{\mathcal{H}}_m(i) = \left[\begin{array}{c} {\bf h}_{k,m}^1 \\
{\bf h}_{k,m}^2\end{array}\right], \boldsymbol{\eta}_k(i) = \left[\begin{array}{c} \boldsymbol{\eta}_{k,1}(2i-1) \\
\boldsymbol{\eta}_{k,2}(2i)\end{array}\right],
\end{equation}
\begin{equation}
{\bf n}(i) = \left[\begin{array}{c} {\bf n}_1(2i-1) \\
{\bf n}_2(2i)\end{array}\right]
\end{equation}

The $2M \times 1$ received vectors ${\bf y}_m(i)$ are linearly
combined with the $2M \times 2$ parameter matrix ${\bf W}_{k,m}$ of
the desired user $k$ of the $m$th antenna at the receiver end to
provide the soft estimates
\begin{equation}
{\bf z}_m(i)= {\bf W}^{H}_{k,m}(i){\bf y}_{m}(i)=[z_{k,m}(2i-1),
z_{k,m}(2i)]^T
\end{equation}
By collecting the soft estimates ${\bf z}_m(i)$ at each receive
antenna, the designer can also exploit the spatial diversity at the
receiver as given by
\begin{equation}
{\bf z}(i) = \sum_{m=1}^{N_r} \boldsymbol{\alpha}_{l}(i) {\bf
z}_{l}(i)
\end{equation}
where $\boldsymbol{\alpha}(i) = diag\Big( \alpha_{l,1}(i),~
\alpha_{l,2}(i) \Big)$ are the gains of the combiner at the
receiver, which can be made equal to each other leading to Equal
Gain Combining (ECG) or proportional to the channel parameters in
accordance with Maximal Ratio Combining (MRC) \cite{rappa}.

\section{Space-time linearly constrained receivers based on the CCM design criterion}

Consider the $2M$-dimensional received vector at the $m$th receiver
${\bf y}(i)$, the $2M\times 2L_{p}$ constraint matrices
$\boldsymbol{\mathcal{C}}_k$ and $\bar{\boldsymbol{\mathcal{C}}}_k$
that were defined in (6) and the $2L_{p}\times 1$ space-time channel
vector $\boldsymbol{\mathcal{H}}_m(i)$ with the multipath components
of the unknown channels from Tx1 and Tx2 to the $m$th antenna at the
receiver. The space-time linearly constrained receiver design
according to the CCM criterion corresponds to determining an FIR
filter matrix ${\bf W}_{k,m}(i)=\Big[ {\bf w}_{k,m}(i)^T, \bar{\bf
w}_{k,m}(i)^T \Big]^T $ with dimension $2M\times 2$ that provides an
estimate of the desired symbol at the $m$th antenna of the receiver
as given by
\begin{equation}
\hat{\bf b}_{k}(i) = \rm{sgn} \Big( \Re \Big[ {\bf
W}_{k,m}^{H}(i){\bf y}_m(i) \Big]\Big) + j~ \rm{sgn}~ \Big(\Im \Big[
{\bf W}_{k,m}^{H}(i){\bf y}_m(i) \Big]\Big)
\end{equation}
where $\rm{sgn}(\cdot)$ is the signum function, $\Re(.)$ selects the
real component, $\Im(.)$ selects the imaginary component and ${\bf
W}_{k,m}(i)$ is optimized according to the CM cost functions
\begin{equation}
\label{costfunction1} J_{CM}({\bf w}_{k,m}) = E\Big[(|{\bf
w}_{k,m}^{H}(i){\bf y}_m|^{2}-1)^{2}\Big]
\end{equation}
\begin{equation}
\label{costfunction2} J_{CM}(\bar{\bf w}_{k,m}) = E\Big[(|\bar{\bf
w}_{k,m}^{H}(i){\bf y}_m|^{2}-1)^{2}\Big]
\end{equation}
subject to the set of constraints described by
\begin{equation}
\label{constraints} \boldsymbol{\mathcal{C}}_k^{H}{\bf w}_{k,m}(i) =
\nu ~ \boldsymbol{\mathcal{H}}_m(i), ~~~
\bar{\boldsymbol{\mathcal{C}}}_k^{H}{\bf w}_{k,m}(i) = \nu ~
\boldsymbol{\mathcal{H}}_m^*(i)
\end{equation}
where $\nu$ is a constant to ensure the convexity of
(\ref{costfunction1}) and (\ref{costfunction2}). Our approach is to
consider the parameter vector design problem in
(\ref{costfunction1}) and (\ref{costfunction2}) through the
optimization of the two filters ${\bf w}_{k,m}(i)$ and $\bar{\bf
w}_{k,m}(i)$ in a simultaneous fashion. The optimization of each
parameter vector aims to suppress the interference and estimate the
symbol transmitted by a given transmit antenna. The expressions for
the filters of the space-time CCM linear receiver are given by
\begin{equation}
\begin{split}
\label{filter1} {\bf w}_{k,m}(i) & = {\bf R}^{-1}_{k,m}(i)\Big[{\bf
d}_{k,m}(i) - \boldsymbol{\mathcal{C}}_k
(\boldsymbol{\mathcal{C}}_k^{H}{\bf
R}^{-1}_{k,m}(i)\boldsymbol{\mathcal{C}}_k)^{-1}\\ & \quad \times
\Big(\boldsymbol{\mathcal{C}}_k^{H}{\bf R}^{-1}_{k,m}(i) {\bf
d}_{k,m}(i) -\nu~\boldsymbol{\mathcal{H}}_m(i)\Big)\Big]
\end{split}
\end{equation}
\begin{equation}
\begin{split}
\label{filter2} \bar{\bf w}_{k,m}(i) & = \bar{\bf
R}^{-1}_{k,m}(i)\Big[\bar{\bf d}_{k,m}(i) -
\bar{\boldsymbol{\mathcal{C}}_k}
(\bar{\boldsymbol{\mathcal{C}}}_k^{H}\bar{\bf
R}^{-1}_{k,m}(i)\bar{\boldsymbol{\mathcal{C}}}_k)^{-1}\\ & \quad
\times \Big( \bar{\boldsymbol{\mathcal{C}}}_k^{H} \bar{\bf
R}^{-1}_{k,m}(i) \bar{\bf d}_{k,m}(i)
-\nu~\boldsymbol{\mathcal{H}}_m^*(i)\Big)\Big]
\end{split}
\end{equation}
The expressions in (\ref{filter1}) and (\ref{filter2}) are not
closed form as they depend on the previous values of the estimators
and require a complexity $O((2M)^3)$ to invert the matrices. Note
also that (\ref{filter1}) and (\ref{filter2}) assume the knowledge
of the space-time channel parameters. However, in applications where
multipath is present these parameters are not known and thus channel
estimation is required. In the next section, we present a method to
blindly estimate channels exploiting the STBC structure.

\section{Space-time Channel Estimation}

In this section, we present a framework that exploits the signature
sequences of the desired user and the unique structure of STBC for
blind channel estimation. Let us consider the received vector ${\bf
y}_m(i)$ at the $m$th Rx, its associated $2M \times 2M$ covariance
matrix ${\bf R}_m=E[{\bf y}_m(i){\bf y}_m^H(i)]$, the space-time $2M
\times 2L_{p}$ constraint matrices $\boldsymbol{\mathcal{C}}_k$ and
$\bar{\boldsymbol{\mathcal{ C}}}_k$ given in (6), the space-time
channel vector $\boldsymbol{\mathcal{H}}_m(i)$ and from (5) we have
that the $k$th user space-time coded received signals without ISI
and noise are given by
\begin{equation}
{\bf x}_k(i)=A_k b_k(2i-1)\boldsymbol{\mathcal{C}}_k
\boldsymbol{\mathcal{H}}_m(i), ~~ \bar{\bf x}_k(i)=A_k
b_k(2i)\bar{\boldsymbol{\mathcal{C}}}_k
\boldsymbol{\mathcal{H}}_m^*(i)
\end{equation}
Let us perform singular value decomposition (SVD) on the space-time
$JM \times JM$ covariance matrix ${\bf R}_m$. By neglecting the ISI,
we have:
\begin{equation}
\begin{split}
{\bf R}_m & = \sum_{k=1}^{K}E[{\bf x}_{k}{\bf x}_{k}^{H}] +
E[\bar{\bf x}_{k}\bar{\bf x}_{k}^{H}] + \sigma^{2}{\bf I} \\ & =
[{\bf V}_{s}~ {\bf V}_{n}] \left[\begin{array}{c c}
\boldsymbol{\Lambda}_{s} + \sigma^{2} {\bf I} & {\bf 0} \\ {\bf 0} &
\sigma^{2} {\bf I}
\end{array}\right] [{\bf V}_{s} ~{\bf V}_{n}]^{H}
\end{split}
\end{equation}
where ${\bf V}_{s}$ and ${\bf V}_{n}$ are the signal and noise
subspaces, respectively. Since the signal and noise subspaces are
orthogonal, we have the following conditions
\begin{equation}
{\bf V}_{n}^{H} {\bf x}_{k}(i)={\bf V}_{n}^{H}
\boldsymbol{\mathcal{C}}_k \boldsymbol{\mathcal{H}}_m(i) = {\bf 0}
\end{equation}
\begin{equation}
{\bf V}_{n}^{H}\bar{\bf x}_{k}(i)={\bf V}_{n}^{H}
\bar{\boldsymbol{\mathcal{C}}}_k \boldsymbol{\mathcal{H}}_m^*(i) =
{\bf 0}
\end{equation}
and hence we have
\begin{equation}
\begin{split}
\boldsymbol{\Omega} =
\boldsymbol{\mathcal{H}}_m(i)^{H}\boldsymbol{{\mathcal
C}_{k}}^{H}{\bf V}_{n}{\bf V}_{n}^{H} \boldsymbol{\mathcal{C}}_k
\boldsymbol{\mathcal{H}}_m(i)={ 0}
\end{split}
\end{equation}
\begin{equation}
\begin{split}
\bar{\boldsymbol{\Omega}} =
\boldsymbol{\mathcal{H}}_m^{T}(i)\bar{\boldsymbol{\mathcal{C}}}_k^{H}{\bf
V}_{n} {\bf V}_{n}^{H} \bar{\boldsymbol{\mathcal{C}}}_k
\boldsymbol{\mathcal{H}}_m^*(i)={ 0}
\end{split}
\end{equation}
From the conditions above and taking into account the conjugate
symmetric properties induced by STBC \cite{Li}, it suffices to
consider only $\boldsymbol{\Omega}$, which allows the recovery of
$\boldsymbol{\mathcal{H}}_m(i)$ as the eigenvector corresponding to
the smallest eigenvalue of the matrix $\boldsymbol{{\mathcal
C}_{k}}^{H}{\bf V}_{n}{\bf V}_{n}^{H} \boldsymbol{\mathcal{C}}_k$,
provided ${\bf V}_{n}$ is known. In this regard,
$\boldsymbol{\mathcal{H}}_m(i)$ belongs to the null space of
$\boldsymbol{{\mathcal C}_{k}}^{H}{\bf V}_{n}{\bf V}_{n}^{H}
\boldsymbol{\mathcal{C}}_k$ and is a linear combination of all
eigenvectors corresponding to eigenvalue zero. If
$\boldsymbol{{\mathcal C}_{k}}^{H}{\bf V}_{n}{\bf V}_{n}^{H}
\boldsymbol{\mathcal{C}}_k$ has multiple zero-valued eigenvalues, we
choose the eigenvalue with smallest index as the solution. To avoid
the SVD on ${\bf R}_m$ and overcome the need for determining the
noise subspace rank that is necessary to obtain ${\bf V}_{n}$, we
resort to the following approach.

\textit{Lemma:} Consider the SVD on ${\bf R}_m$ as in (18), then we
have:
\begin{equation}
\lim_{p \rightarrow \infty} ({\bf R}_m/\sigma^{2})^{-p} = {\bf
V}_{n} {\bf V}_{n}^{H}
\end{equation}
\textit{Proof:} Using the decomposition in (18) and since ${\bf I} +
\boldsymbol{\Lambda}_{s}/\sigma^2$ is a diagonal matrix with
elements strictly greater than unity, we have the following limit as
$p\rightarrow \infty$:
\begin{equation}
\begin{split}
({\bf R}_m/\sigma^{2})^{-p} & = [{\bf V}_{s}~ {\bf V}_{n}]
\left[\begin{array}{c c} ({\bf I} +
\boldsymbol{\Lambda}_{s}/\sigma^2)^{-p} & {\bf 0} \\
{\bf 0} &  {\bf I}
\end{array}\right] [{\bf V}_{s} ~{\bf V}_{n}]^{H} \\
 & \rightarrow \quad  [{\bf V}_{s}~ {\bf V}_{n}]
\left[\begin{array}{c c} {\bf 0} & {\bf 0} \\
{\bf 0} & {\bf I}
\end{array}\right] [{\bf V}_{s} ~{\bf V}_{n}]^{H} = {\bf V}_{n}{\bf
V}_{n}^{H}.
\end{split}
\end{equation}
To blindly estimate the space-time channel of user $k$ at the $m$th
antenna of the receiver we propose the following optimization:
\begin{equation}
{\hat{\boldsymbol{{\mathcal H}}}}_{m}(i) = \arg
\min_{{{\boldsymbol{{\mathcal H}}}}}~~{{{\boldsymbol{{\mathcal
H}}}}^{H} \boldsymbol{{\mathcal C}_{k}}^{H}\hat{\bf
R}^{-p}_m(i)\boldsymbol{{\mathcal C}_{k}}{{\boldsymbol{{\mathcal
H}}}}}
\end{equation}
subject to { $||{\hat{\boldsymbol{{\mathcal H}}}}_{m}(i)||=1$},
where $p$ is an integer, $\hat{\bf R}_m(i)$ is an estimate of the
covariance matrix ${\bf R}_m(i)$ and whose solution is the
eigenvector corresponding to the minimum eigenvalue of the
$JL_{p}\times JL_{p}$ matrix $\boldsymbol{{\mathcal
C}_{k}}^{H}\hat{\bf R}^{-p}_m(i) \boldsymbol{{\mathcal C}_{k}}$ that
can be obtained using SVD. The performance of the estimator can be
improved by increasing $p$ even though our studies reveal that it
suffices to use powers up to $p=2$ to obtain a good estimate of
${\bf V}_{n}{\bf V}_{n}^{H}$. For the space-time block coded CCM
receiver design, we employ the matrix ${\bf R}_{k,m}$ instead of
${\bf R}_m$ to avoid the estimation of both ${\bf R}_m$ and ${\bf
R}_{k,m}$, and which shows no performance loss as verified in our
studies. The computational complexity of the proposed space-time
estimator in (25) is $O((2L_p)^3)$.

\section{Blind Adaptive SG Algorithms for Receiver and Channel Parameter Estimation}

Here we describe SG algorithms for the blind estimation of the
channel and the parameter vector ${\bf w}_{k,m}$ of the proposed
space-time linear receivers using the CCM criterion.

\subsection{Space-Time Constrained Constant Modulus SG Algorithm}

To derive an CCM-SG algorithm let us transform the constrained
optimization problem given by (12)-(14) into unconstrained problems
in the form of the Lagrangians
\begin{equation}
{\mathcal{L}}({\bf w}_{k,m})  = (|z_{k,m}[i]|^{2} - 1)^{2} +
2\Re\Big[({\boldsymbol{{\mathcal C}}_{k}^{H}{\bf w}_{k,m}[i]-\nu
\boldsymbol{{{\mathcal H}}}}_{m}[i])^{H} \boldsymbol{\lambda}_1\Big]
\end{equation}
\begin{equation}
{\mathcal{L}}(\bar{\bf w}_{k,m})  = (|{\bar z}_{k,m}[i]|^{2} -
1)^{2} + 2\Re\Big[({\bar{\boldsymbol{{\mathcal C}}}_{k}^{H}\bar{\bf
w}_{k,m}[i]-\nu \boldsymbol{{{\mathcal H}}}}_{m}^*[i])^{H}
\boldsymbol{\lambda}_2\Big]
\end{equation}
where $z_{k,m}[i] = {\bf w}_{k,m}^{H}[i]{\bf y}_{m}[i]$, ${\bar
z}_{k,m}[i] = \bar{\bf w}_{k,m}^{H}[i]{\bf y}_{m}[i]$ and
$\boldsymbol{\lambda}_1$ and $\boldsymbol{\lambda}_2$ are vectors of
Lagrange multipliers. SG solutions to (26) and (27) can be obtained
by taking the gradient terms of (26) and (27) with respect to ${\bf
w}_{k,m}(i)$ and $\bar{\bf w}_{k,m}(i)$ which yields the following
parameter estimators:
\begin{equation}
\begin{split}
{\bf w}_{k,m}[i+1] & = \boldsymbol{\Pi}_{k}({\bf w}_{k,m}[i] -
\mu_{w}e_{k,m}(i)z_{k,m}^{*}[i]) \\ & \quad +{\bf C}_{k} ({\bf
C}_{k}^{H}{\bf C}_{k})^{-1} \boldsymbol{{\mathcal H}}_{m}[i]
\end{split}
\end{equation}
\begin{equation}
\begin{split}
\bar{\bf w}_{k,m}[i+1] & = \bar{\boldsymbol{\Pi}}_{k}(\bar{\bf
w}_{k,m}[i] - \mu_{w}{\bar e}_{k,m}(i){\bar z}_{k,m}^{*}[i]) \\ &
\quad +\bar{\bf C}_{k} (\bar{\bf C}_{k}^{H}\bar{\bf C}_{k})^{-1}
\bar{\boldsymbol{{\mathcal H}}}_{m}^*[i]
\end{split}
\end{equation}
where $e_{k,m}[i] = (|z_{k,m}[i]|^{2} -1)$, ${\bar e}_{k,m}[i] =
(|{\bar z}_{k,m}[i]|^{2} -1)$, $\boldsymbol{\Pi}_{k} = {\bf I}
-\boldsymbol{{\mathcal C}}_{k}(\boldsymbol{{\mathcal
C}}_{k}^{H}\boldsymbol{{\mathcal C}}_{k})^{-1} \boldsymbol{{\mathcal
C}}_{k}^{H}$, $\bar{\boldsymbol{\Pi}}_{k} = {\bf I} -\bar
{\boldsymbol{\mathcal C}}_{k}(\bar {\boldsymbol{{\mathcal
C}}}_{k}^{H}\bar {\boldsymbol{{\mathcal C}}}_{k})^{-1} \bar
{\boldsymbol{\mathcal C}}_{k}^{H}$. The proposed SG algorithm has
complexity $O(4ML_p)$ in comparison with that of $O((2M)^3)$, as
required by the expressions in (15) and (16).

\subsection{Blind Space-Time SG Channel Estimation Algorithm}

In order to estimate the space-time channel and avoid the SVD on
$\boldsymbol{{\mathcal C}}_{k}^{H}{\bf R}^{-1}_{k}[i]
\boldsymbol{{\mathcal C}}_{k}$, we compute the estimates
$\boldsymbol{{ \Omega}}_{m}[i]={\bf
C}_{k}^{H}\boldsymbol{\hat{\Psi}}_{k}[i]$, where
$\boldsymbol{\hat{\Psi}}_{k}[i]$ is an estimate of the matrix ${\bf
R}^{-1}_{m}[i] {\bf C}_{k}$. The estimate
$\boldsymbol{\hat{\Psi}}_{k}[i]$ is obtained with the following
recursion:
\begin{equation}
\boldsymbol{\hat{\Psi}}_{m}[i] = \alpha
\boldsymbol{\hat{\Psi}}_{m}[i-1] +
\mu_{h}\Big(\boldsymbol{\hat{\Psi}}_{k}[i-1] - {\bf y}_m[i]{\bf
y}_m^{H}[i] \boldsymbol{\hat{\Psi}}_{m}[i-1]\Big)
\end{equation}
where $\boldsymbol{\hat{\Psi}}_{m}(0)=\boldsymbol{{\mathcal C}}_{k}$
and $0<\alpha<1 $. To estimate the space-time channel and avoid the
SVD on $\boldsymbol{{\mathcal C}}_{k}^{H}{\bf R}^{-1}_{k}[i]
\boldsymbol{{\mathcal C}}_{k}$, we employ the variant of the power
method introduced in \cite{douko}
\begin{equation}
{\hat{\boldsymbol{\mathcal H}}}_m (i)= ({\bf I} - \gamma(i)
\boldsymbol{{ \Omega}}_{m}[i]) {\hat{\boldsymbol{\mathcal
H}}}_m(i-1)
\end{equation}
where $\gamma(i)=1/tr[{\boldsymbol{\Gamma}}_{k,m}(i)]$ and we make
$\hat{\boldsymbol{\mathcal H}}_{m}(i)\leftarrow
\hat{\boldsymbol{\mathcal H}}_{m}(i)/||\hat{\boldsymbol{\mathcal
H}}_{m}(i)||$ to normalize the channel. The proposed space-time
channel estimation algorithm has complexity $O(4ML_p)$ and shows
excellent performance.

\section{Simulations}

We evaluate the bit error rate (BER) performance of the proposed
blind space-time block coded receivers based on the CCM design, the
proposed space-time channel estimation method in terms of mean
squared error (MSE) performance and the SG adaptive algorithms. We
also compare the proposed space-time CCM linear receivers and blind
channel estimation algorithms with some previously reported
techniques, namely, the constrained minimum variance (CMV) with a
single antenna \cite{xu&tsa} and with STBC \cite{Li} and the
subspace receiver of Wang and Poor without \cite{wang&poor} and with
STBC \cite{reynolds}. The DS-CDMA system employs randomly generated
spreading sequences of length $N=32$, employs one or two transmit
antennas with the Alamouti STBC \cite{alamouti} and employs one
receive antenna with MRC. All downlink channels assume that
$L_{p}=6$ as an upper bound. We use three-path channels with
relative powers $p_{l,m}^{1,2}$ given by $0$, $-3$ and $-6$ dB, with
path-spacing given by a discrete uniform random variable between $1$
and $2$ chips. The sequence of channel coefficients for each
transmit antenna $n_t=1,2$ and each receive antenna $m=1,2$ is
$h_{l,m}^{n_t}(i)=p_{l,m}^{n_t} \alpha_{l,m}^{n_t}(i)$
($l=0,1,2,~\ldots$), where $\alpha_{l,m}^{n_t}(i)$, is obtained with
Clarke's model \cite{rappa}.

In the first scenarios, shown in Figs. 2 and 3, we evaluate the BER
convergence performance of the proposed SG algorithms for both
receiver and channel parameter estimation. We consider a system with
$8$ users, the power level distribution among the interferers
follows a log-normal distribution with associated standard deviation
of $3$ dB and the desired user power level corresponds to the
signal-to-noise-ratio (SNR) defined by $SNR=E_{b}/N_{0}=15$ dB.
After $1500$ symbols, $6$ additional users enter the system and the
power level distribution among interferers is loosen with associated
standard deviation being increased to $6$ dB. The results, shown in
Fig. 2, indicate that the proposed CCM-based SG algorithms
outperform the existing blind algorithms (CMV-based and Subspace)
and approach the performance of supervised algorithms. The results
for channel estimation, depicted in Fig. 3, show that the proposed
STBC-based SG algorithm which equips both CCM-based and CMV-based
schemes outperforms the single-antenna method of \cite{douko} and
approaches the highly complex subspace approach of \cite{reynolds}.

\begin{figure}[!htb]
\begin{center}
\def\epsfsize#1#2{0.825\columnwidth}
\epsfbox{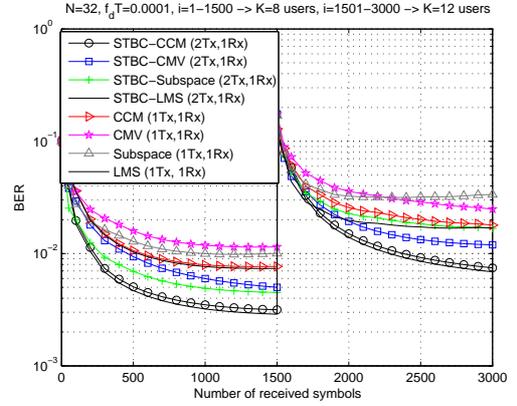} \caption{BER performance versus number of
received symbols .}
\end{center}
\end{figure}

\begin{figure}[!htb]
\begin{center}
\def\epsfsize#1#2{0.825\columnwidth}
\epsfbox{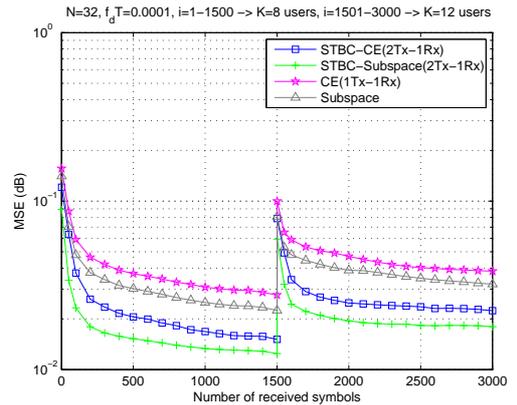} \caption{MSE channel estimation performance
versus number of symbols .}
\end{center}
\end{figure}

The BER performance versus SNR and number of users is illustrated in
Fig. 4. In these scenarios, we considered data packets of $P=1500$
symbols, $2$ transmit antennas, $1$ receive antenna and measured the
BER after $200$ independent transmissions. The plots indicate that
the proposed STBC-CCM receiver achieves the best performance,
followed by the subspace receiver with STBC of \cite{reynolds} and
the STBC-CMV. A substantial capacity increase and performance
improvement is verified for the schemes with multiple antennas.

\begin{figure}[!htb]
\begin{center}
\def\epsfsize#1#2{0.825\columnwidth}
\epsfbox{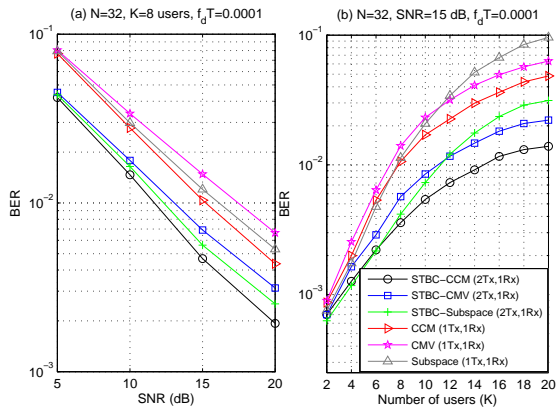} \caption{BER performance versus (a) $E_b/N_0$ and
(b) number of users.}
\end{center}
\end{figure}

\section{Conclusions}

We have presented in this work blind adaptive linear multiuser
receivers for DS-CDMA systems using multiple transmit antennas and
space-time block codes (STBC) in multipath channels. We considered a
CCM design criterion based on constrained optimization techniques
and described low-complexity SG adaptive algorithms for estimating
the parameters of the linear receivers. The receiver was designed in
order to exploit the unique structure imposed by both spreading
codes and STBC. We also developed a blind space-time channel
estimation scheme for STBC systems based on the subspace approach
along with an efficient SG algorithm for channel estimation.
Numerical results for a downlink scenario have shown that the
proposed techniques outperform existing schemes in realistic
scenarios.

\end{document}